\documentclass{article}
\usepackage{spconf,amsmath,graphicx}
\usepackage{amsmath}
\usepackage{amsxtra}
\usepackage{graphicx}
\usepackage{epsfig}
\usepackage{latexsym}
\usepackage{amsfonts}
\usepackage{rawfonts}
\usepackage[latin1]{inputenc}
\usepackage[T1]{fontenc}
\usepackage{calc}
\usepackage{url}
\usepackage{enumerate}
\usepackage{color}
\usepackage{amssymb}
\usepackage{upref}
\usepackage{epic,eepic}
\usepackage{times}
\usepackage{cite}
\usepackage{dsfont}
\usepackage{mathrsfs}
\usepackage{verbatim}

\DeclareMathOperator*{\argmax}{argmax}
\newtheorem{definition}{\bf Definition}
\newtheorem{lemma}{\bf Lemma}

\newlength{\aligntop}
\newlength{\alignbot}
\makeatletter
\renewenvironment{align}{%
  \vspace{\aligntop}
  \start@align\@ne\st@rredfalse\m@ne
}{%
  \math@cr \black@\totwidth@
  \egroup
  \ifingather@
    \restorealignstate@
    \egroup
    \nonumber
    \ifnum0=`{\fi\iffalse}\fi
  \else
    $$%
  \fi
  \ignorespacesafterend%
  \vspace{\alignbot}\par\noindent
}

\title{Matching Theory for Priority-Based Cell Association in the Downlink of Wireless Small Cell Networks\vspace{-0.1cm}}
\name{Omid Semiari$^\dag$, Walid Saad$^\dag$, Stefan Valentin$^\ddag$, Mehdi Bennis$^\ast$, Behrouz Maham$^{\star}$ 
\vspace{-0.4cm}}
\address{\small $^\dag$ECE Department, University of Miami, Coral Gables, FL, USA, Email: \url{o.semiari@umiami.edu}, \url{walid@miami.edu}\\
\small $^\ddag$Bell Labs, Alcatel-Lucent, Germany, Email: \url{stefan.valentin@alcatel-lucent.com}\\
\small $^\ast$Centre for Wireless Communications-CWC, University of Oulu, Finland, Email: \url{bennis@ee.oulu.fi}\\
\small $^\star$School of Electrical and Computer Engineering, University of Tehran, Iran, Email: \url{bmaham@ut.ac.ir}\vspace{-0.6cm}
\thanks{This research is supported by the National Science Foundation under Grants CNS-1253731 and CNS-1406947.}}

\begin{document}
\ninept
\maketitle
\vspace{-5cm}
\begin{abstract}
The deployment of small cells, overlaid on existing cellular infrastructure, is seen as a key feature in next-generation cellular systems. In this paper,  the problem of user association in the downlink of small cell networks (SCNs) is considered. The problem is formulated as a many-to-one matching game in which the users and SCBSs rank one another based on utility functions that account for both the achievable performance, in terms of rate and fairness to cell edge users, as captured by newly proposed priorities. To solve this game, a novel distributed algorithm that can reach a stable matching is proposed. Simulation results show that the proposed approach yields an average utility gain of up to $65\%$ compared to a common association algorithm that is based on received signal strength. Compared to the classical deferred acceptance algorithm, the results also show a $40\%$ utility gain and a more fair utility distribution among the users.
\end{abstract}
\begin{keywords}
Small cell networks; Matching theory; Cell association.\vspace{-0.5cm}
\end{keywords}
\section{Introduction}\vspace{-0.3cm}
\label{sec:intro}
Smartphones have significantly increased the traffic load in current cellular networks and this trend is expected to continue in the next few years\cite{17}. Meeting the demand generated by this increasing traffic requires significant changes to current cellular architecture. One promising approach to address this problem is via the concept of small cell networks (SCNs)\cite{1,2}. SCNs allow to improve the capacity and coverage of wireless networks by reducing the distance between users and their serving base stations. This is done by deploying small cell base stations (SCBSs), overlaid on current macrocell networks and connecting to existing backhauls such as DSL\cite{14}.

The deployment of small cells introduces numerous challenges in terms of interference management, resource allocation, and network modeling\cite{17,1,2,14,3,4,8,5,6,7,9,10}. In particular, cell association is an important challenge in SCNs. For instance, directly deploying classical macrocell-oriented cell association schemes in SCNs can lead to inefficient association due to the factors such as heterogeneous capabilities and varying coverage areas~\cite{3}. In \cite{4,8}, the authors proposed several biased cell association approaches, in which the SCBSs' coverage areas are increased, to improve the network's overall rate by associating more user equipments~(UEs) to SCBSs. Nevertheless, one practical limitation of biasing is the use of overhead channels shared by all SCBSs. Thus, different interference cancelation and power control algorithms have been proposed in \cite{5,6,7,9} to address this problem.
A new dynamic cell association approach to maximize sum rate is introduced in \cite{10} allowing UEs to adopt a heuristic cell range expansion algorithm for load balancing. However, depending on the bias value, this method may cause certain UEs to suffer from signal-to-interference plus-noise ratio (SINR) degradation.

Although these works provide interesting insights on cell association, they are not user-centric and are mostly based on signal strength or SINR. Indeed, most of these existing works require network-level coordination which increases both complexity and overhead and is undesirable in SCNs which are expected to be self-organizing. One prospective approach to providing self-organizing cell association in SCNs is via the powerful tools of \emph{matching theory}\cite{15}. While matching theory has recently attracted a lot of attention in wireless networks, such as for associating channels in ad hoc and cognitive networks \cite{11,12}, most of these works only focus on the maximization of SINR-based utilities and do not handle SCN-specific challenges. Moreover, these approaches do not offer satisfactory solution for non-uniform user distributions and are often unfair to cell-edge users.

The main contribution of this paper is to develop a novel approach for cell association in which users are smartly prioritized based on their location and proximity to the small cells. The problem is formulated as a matching game in which users and base stations~(BSs) rank one another using preferences based on well-defined utility functions. The proposed utilities at each BS capture not only the rates it can offer to users, but also the preference of each user to be associated to other BSs. These utilities also incorporate a new prioritization technique that allows cell-edge UEs to more actively participate in cell association. For solving the game, we propose a novel algorithm based on the deferred acceptance (DA) mechanism. Using this algorithm, we show that the user-cell association problem can reach a stable matching. Simulation results show that the proposed approach gives a considerable gain over both conventional DA \cite{11} and received signal strength indicator (RSSI) approaches\cite{14}. The results also show that the proposed priority-based deferred acceptance algorithm improves the utility distribution among users and increases the average utility of the network. 

The rest of the paper is organized as follows. Section 2 describes the system model. Section 3 defines the problem as the matching game and presents the proposed algorithm. Simulation results are analyzed in Section 4 and conclusions are drawn in Section 5.\vspace{-0.45cm}
\section{System Model}\vspace{-0.35cm}
\label{sec:System Model}
Consider the downlink of an OFDMA SCN having a single macrocell overlaid with $L-1$ SCBSs randomly distributed in the coverage area of the macrocell base station (MBS). We consider an open access scheme in which all UEs are allowed to connect to their preferred tier. We assume that all tiers use the same spectrum, i.e. co-channel deployment~\cite{16}. The total bandwidth $B$ is divided into $N$ subcarriers in the set $\mathcal{N}$ and there are a total of $M$ active users with $\mathcal{M}$ being the set of all users. Hereafter, we use the term ``BS'' to denote either an MBS or an SCBSs in $\mathcal{L}$. The Shannon's achievable capacity of UE $m$ from BS $l$ over subcarrier $j$ is:
\begin{align}\label{eq:1}
\Phi_{ljm}(\gamma_{ljm})=w_{lj}\log(1+\gamma_{ljm}),
\end{align}
where $w_{lj}$ and $\gamma_{ljm}$ denote the bandwidth of subcarrier $j$ and the SINR, respectively.

One important challenge in such an SCN is the problem of associating the UEs to their serving BS. In a conventional setting, each active UE is served by the BS which offers the highest RSSI. From the network's perspective, the cell association is often defined as an optimization problem in which UEs are assigned to BSs ($\mu:\mathcal{M}\to\mathcal{L}$) such that the overall sum utility of the network is maximized:
\begin{align}\label{eq:opt}
\argmax_{\mu} &\sum_{l\in \mathcal{L}}\sum_{m\in\mathcal{M}_{l}}\sum_{j\in\mathcal{N}} \Phi_{ljm}(\gamma_{ljm}),\\
&\text{s.t.,} \,\,\,\,\forall m: \sum_{j\in\mathcal{N}}\Phi_{ljm}(\gamma_{ljm})>\Phi_{\text{th},m},
\end{align}
where $\mathcal{M}_{l}$ denotes a set of all users associated to BS $l$. $\Phi_{\text{th},m}$ represents the capacity threshold determined by the quality of service (QoS) requirements of UE $m$. The problem given by (\ref{eq:opt}) is known to be NP-hard and complex to solve, due to non-linear and combinatorial nature of the assignment problem~\cite{13}.


In SCNs, it is desirable to develop a self-organizing cell association solution due to the network scale, the unplanned deployment of SCBSs, and the limited SCBS coordination due to the finite-capacity backhaul~\cite{14}. Hence, new approaches for cell association are needed. One promising approach is via matching theory, as discussed next.\vspace{-0.2cm}

\section{Cell Association As a Matching Game}\vspace{-0.1cm}
\label{sec:Cell Association Matching}
A matching game is defined by two sets of players that evaluate one another using well-defined preference relations\cite{15}. We formulate the proposed cell association problem in SCNs as a many-to-one matching game in which a set of users $\mathcal{M}$ will be assigned to a set of BSs $\mathcal{L}$, where each UE will be assigned to at most one BS. We assume that an arbitrary BS $l$ can serve a maximum number of UEs (quota) $q_{l}$ in the downlink.
Depending on the channel quality or equivalently SINR values, each UE builds a \emph{preference relation} $\succ_{m}$ over subsets of BSs and being unmatched $\emptyset$. In fact, via the transmission of initial ranging signals, each UE $m$ is able to form a $L\times N$ channel matrix $\textbf{H}_{m}$ in which each element $h_{ljm}$ is the channel gain of the subcarrier $j$ used for the link between BS $l$ and user $m$. We will show that we can use these preference relations to obtain performance gain over conventional cell association approaches. Further, each BS has a preference $\succ_{l}$ over the subset of UEs based on a pre-defined utility function. Iteratively, the UEs propose to their most preferred BS according to their preferences and BSs accept or reject proposals based on utilities they assign to their applicants. With this in mind, a matching $\mu$ between SCBSs and users is defined as follows:\vspace{-0.2cm}
\begin{definition} A \emph{matching} is defined as a function from the set $\mathcal{M} \cup \mathcal{L}$ into the set of $\mathcal{M} \cup \mathcal{L}$ such that:
1) $|\mu(m)|=1$ for each user and $\mu(m) \in \mathcal{L}\cup\emptyset$,
2) $|\mu(l)|\le q_{l}$ for BS $l$. Also, $\mu(l) \subseteq \mathcal{M}\cup\emptyset$, and
3) $m \in \mu(l)$ if and only if $\mu(m)=l$.
\end{definition}\vspace{-0.2cm}
Therefore, the tuple $\left(\mathcal{L},\mathcal{M},\succ_{\mathcal{L}},\succ_{\mathcal{M}},\mathcal{Q}\right),$ determines the cell association matching problem with $\succ_{\mathcal{L}}=\left\{\succ_{l}\right\}_{l \in \mathcal{L}}$ being the preference set of the BSs, $\succ_{\mathcal{M}}=\left\{\succ_{m}\right\}_{m \in \mathcal{M}}$ being the preference set of the users, and $\mathcal{Q}=\{q_{l}|\,\, \forall l \in \mathcal{L}\}$ being the BSs' quota vector.\vspace{-0.3cm}
\subsection{Priority-based Preferences}\vspace{-0.2cm}
 \begin{figure}
  \centering
  \centerline{\includegraphics[width=8cm]{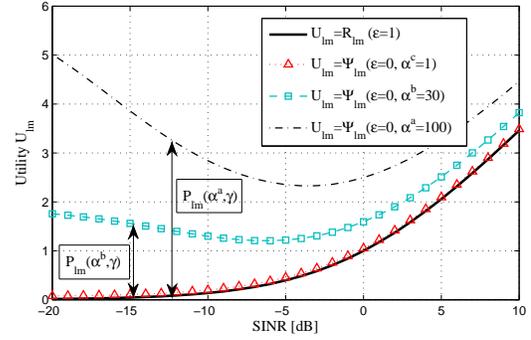}}
\vspace{-0.3cm}
  \caption{Utility function at BS side versus SINR and for different priorities. Here, $\zeta_1=0.1$ and $\zeta_2=3$ and $\alpha_m \in \{100, 30, 1\}$.}
\label{fig:exam}\vspace{-0.5cm}
\end{figure}
To fully describe the matching $\mu$, next, we define the preferences by each side of the game.\vspace{-0.4cm}
\subsubsection{Users' Preferences}\vspace{-0.2cm}
From the users' side, each UE seeks to maximize its own, individual utility function. Therefore, from the UEs' point of view, we use rates as the utility functions.
Thus, using the estimated channel coefficient matrix $\textbf{H}_{m}$, user $m$ shapes its $L\times N$ achievable data rate matrix, i.e. $\bf{\Phi}_{\text{$m$}}$, whose elements are defined by (\ref{eq:1}), where
\begin{align}\label{eq:2}
\gamma_{ljm}=\frac{p_{lj}h_{ljm}}{\sum_{k=1, k\neq l}^{L} p_{kj}h_{kjm}+\sigma^{2}}.
\end{align}
(\ref{eq:2}) represents the potential achievable SINR for UE $m$ from BS $l$ at subcarrier $j$. $p_{lj}$ and $\sigma^{2}$ denote the transmit power of BS $l$ over subcarrier $j$ and the variance of the receiver's Gaussian noise, respectively. In order to rank BSs, each user $m$ takes the average of the achievable data rates from each BS over all $N$ subcarriers. From (1) and (4), the $1 \times L$ utility vector of user $m$, $\text{R}_{m}$, is given by:
\begin{align}\label{eq:5}
R_{lm}(\gamma_{ljm})\!\!=\!\!\frac{1}{N}\!\sum_{j=1}^{N}\Phi_{ljm}(\gamma_{ljm})\!=\!\!\frac{1}{N}\!\sum_{j=1}^{N}w_{lj}\log\!\! \left(\!1+\gamma_{ljm}\right),
\end{align}
where $R_{lm}$ denotes the $l$-th element of $\text{R}_{m}$, that is the average achievable rate for user $m$ from BS $l$ over $N$ subcarriers.
A BS $l$ is said to be \emph{acceptable} for user $m$, i.e. BS $l\succ_{m}\emptyset$  if and only if $R_{lm}>\Phi_{\text{th}}$. In addition, BS $t\succ_{m}$ BS $s$, if and only if $R_{tm}>R_{sm}$. Thus, the \emph{preference matrix} of users, $\mathbf{M}_{M \times L}$, can be obtained whose $m$-th row, $\chi_m=\mathbf{M}(m,:)$, is the preference vector of the user $m$. This vector is a subset of $\mathcal{L}$ that is sorted in descending order based on the utility vector $\text{R}_{m}$.\vspace{-0.4cm}
\subsubsection{Preferences of the MBS and SCBSs}\label{sec:priority}\vspace{-0.2cm}
The proposed matching game can be fully represented once the preference of each BS over users is defined. Here, we define a \textit{novel scheme} at the BS side of the game, which gives priority to UEs based on the information gathered by each BS on the UEs' preferences. Most of the matching approaches in the literature focus on the utilities that only depend on SINR information\cite{11,12}. We show that utilizing the information concealed in the UEs' preferences offers considerable gains in rates and other metrics of the network. Therefore, unlike prior works, we propose novel utilities that depend on such information. Suppose that users send their preference vector to each BS they wish to associate with. Hence, each BS can form a \emph{chance} vector for its UE applicants, $\text{C}_{1 \times M}$, whose elements are chosen from $\{0,1\}$. Then, the BS assigns priorities to its UE applicants based on this vector. Depending on the type of priority given to a UE applicant, BS will promote the utility of that particular UE. If UE applicant $m$ has another option to apply to according to its preference vector, the BS sets $\text{C}(m)=1$, otherwise $\text{C}(m)=0$. Consequently, chance vectors are different at each BS and get updated for each set of new applicants. Thus, instead of ranking users by only rate maximization criterion, each BS takes the chance of each user into account.
\begin{figure}[!t]
  \begin{center}
   \vspace{-0.2cm}
    \includegraphics[width=8cm]{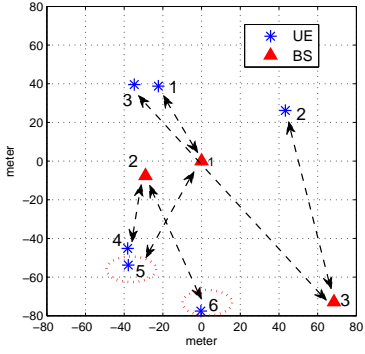}
    \vspace{-0.3cm}
    \caption{\label{fig:snap} An example of cell association with PDA. The black dotted lines show the matching among randomly distributed UEs and BSs.}
  \end{center}\vspace{-1cm}
\end{figure}

Next, we describe the matching approach at the BSs side while clarifying the user assignment procedure. Using (\ref{eq:5}), the utility function of each BS is defined as follows:
\begin{align}
U_{lm}(\alpha_m, \epsilon, \gamma_{ljm})=\epsilon R_{lm}(\gamma_{ljm})+\left(1-\epsilon\right)\Psi_{lm}(\alpha_m,\gamma_{ljm}),
\end{align}
where $U_{lm}$ denotes the utility of user $m$ given by BS $l$ which is a function of priority coefficients $\alpha_m$, resemblance factor $\epsilon\in\{0,1\}$, and $\gamma_{ljm}$. Hence, $\text{UE}_{m}\succ_{l}\text{UE}_{m'}$ if and only if $U_{lm}>U_{lm'}$. Clearly, UE $m$ will be rejected if its utility $U_{lm}$ is not one of the $q_l$ highest utilities. If two users are identified with the same priority by the BS they apply to, then $\epsilon=1$. Otherwise, the BS assigns $\epsilon=0$ to the utility of those two users. Function $\Psi_{lm}(\alpha_m,\gamma_{ljm})$ in (6) is given by:
\begin{align}
&\Psi_{lm}(\alpha_m,\gamma_{ljm})=P_{lm}(\alpha_m,\gamma_{ljm})+R_{lm}(\gamma_{ljm})\\\nonumber
&=\frac{1}{N}\sum_{j=1}^{N}w_{lj}\left(\frac{\alpha_m\zeta_1 }{\log\left(\zeta_2+\alpha_m\gamma_{ljm}\right)}+\log \left(1+\gamma_{ljm}\right)\right).
\end{align}
The \emph{promotion function} $P_{lm}(\alpha_m,\gamma_{ljm})$ represents the amount of promotion given to each class of users. That is, a BS increases the value of each user's achievable rate, based on the user's priority $\alpha_m$. The higher the priority that a certain user has, the more promotion it will receive from the BS. Basically, we let $\alpha_m \in \{\alpha^a,\alpha^b,\alpha^c\}$ indicate the first, second and third priority coefficients, respectively. The constant parameters $\zeta_1$ and $\zeta_2$ are used to control the shape of $\Psi(\alpha_m,\gamma_{ljm})$. Fig.~\ref{fig:exam} illustrates how each type of priority impacts the utility function $U_{lm}$. The parameter $\epsilon$ is used to avoid prioritizing two users that have the same priority, since the promotion is a function of SINR. Clearly, the proposed priorities allow to "promote" users that are experiencing a relatively low SINR, thus allowing them to have a better BS association. Following describes the prioritizing procedure.

Once the UE proposals are sent to an arbitrary BS $l$, applicants of that BS can be divided into three groups of priorities as follows:\\
\textbf{1st Priority:} This includes UEs who have BS $l$ as both their first and their only remaining preference. Therefore, these applicants have been accepted by BS $l$ in the first iteration of proposals. That is, all $m$ who
$\text{C}(m)=0\,\,\,\text{and} \,\,\,\chi_{m}(1)=l.$\\
\textbf{2nd Priority:} This includes users for whom BS $l$ is not the first preference but it is the only remaining BS in the preference list. In other words,
 $\text{C}(m)=0\,\,\,\text{and} \,\,\,\chi_{m}(1)\ne l.$\\
\textbf{3rd Priority:} This includes the users that, if and when rejected by BS $l$, they still have other choices in their preference list, i.e.,
$\exists\, l' \!\in \mathcal{L}\setminus l:\,\,\,\,\,  R_{l}>R_{l'}>\Phi_{\text{th}},$
$\text{or equivalently}\, \text{C}(m)=1$.

These priorities are defined such that no UE can belong to two different priority groups. We will show that this scheme will increase the overall utility and the average rate of the SCBSs with worst-case rates, by having more users involved in the association process.\vspace{-0.1cm}
\begin{figure}[!t]
  \begin{center}
   \vspace{-0.4cm}
    \includegraphics[width=7.5cm]{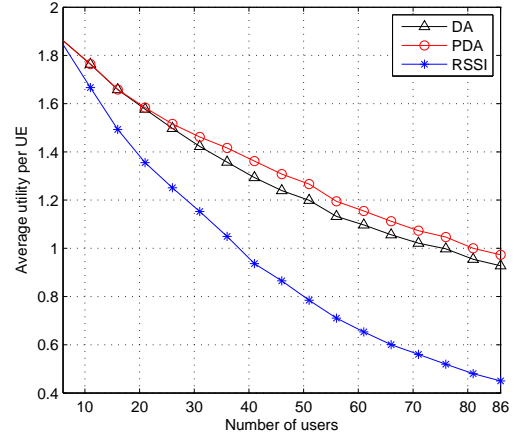}
    \vspace{-0.3cm}
    \caption{\label{fig:fig2} Average utility per UE for PDA, RSSI and DA algorithms.}
  \end{center}\vspace{-0.3cm}
\end{figure}

\begin{figure}[!t]
  \begin{center}
   \vspace{-0.4cm}
    \includegraphics[width=7.5cm]{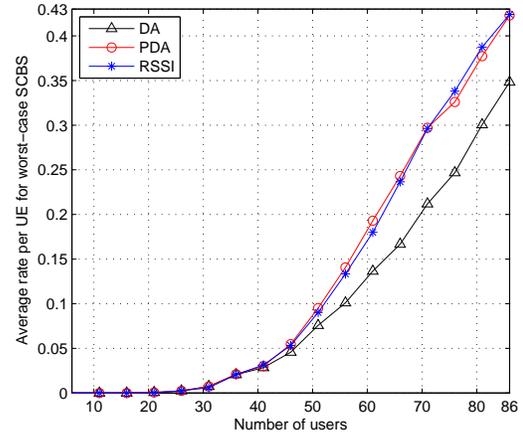}
    \vspace{-0.3cm}
    \caption{\label{fig:fig3} SINR-based average rates of the worst-case SCBS.}
  \end{center}\vspace{-1cm}
\end{figure}
\subsection{Proposed Priority-based Deferred Acceptance Algorithm}\vspace{-0.1cm}
\label{sec:PDA Algorithm}
In Table~\ref{tab:algo}, we show the various stages of the proposed priority-based deferred acceptance (PDA) algorithm which incorporates the prioritizing procedure in Subsection 3.1. For practical implementation, the preference matrix $\textbf{M}$ can be obtained from Reference Signal Received Power (RSRP) signaling\cite{14}. In addition, since users send their preferences to BSs, no knowledge of SCBS distribution is required. Hence, the priority-based approach is feasible for self-organizing SCN implementation.\vspace{-0.2cm}
\begin{definition} A matching $\mu$ is \emph{stable}, if and only if no pair of $\big\{(m,l)|m \in \mathcal{M}, l \in \mathcal{L}\big\}$ blocks the matching. That is,
\begin{align}
\nexists (m,l)\,\,\, \text{s.t.}\,\,\, m\succ_{l}\mu(l) \,\,\,\text{and}\,\,\, l\succ_{m}\mu(m).
\end{align}
\end{definition}\vspace{-0.4cm}
For the proposed algorithm in Table~\ref{tab:algo}, we can state the following:\vspace{-0.2cm}
\begin{lemma}
The proposed PDA algorithm shown in Table~\ref{tab:algo} is guaranteed to converge to a stable matching.\vspace{-0.2cm}
\end{lemma}
This is a direct result of the fact that the proposed algorithm is based on DA, which is shown to always converge to a stable matching~\cite{15}.

Fig.~\ref{fig:snap} shows an example of a small-scale SCN with $M=6$ and $L=3$. Here, the user preference matrix is derived as $\mathbf{M}_{6 \times 3} \!\!\!=\!\!\!
\begin{pmatrix} 1 & 1 & 1 & 1 & 1 & 1  \\ 0 & 3 & 2 & 2 & 2 & 2 \\ 0 & 2 & 3 & 0 & 0 & 0 \end{pmatrix}^{\textbf{T}}$, where the $m-\text{th}$ row indicates the preference list of the user $m$, $\chi_m$, and $(.)^{\textbf{T}}$ is the transpose operation. The DA, RSSI and PDA approaches will lead to the following matchings: $\mu^{\text{DA}} \!\!\! =\!\!\begin{pmatrix} 1 & 3 & 2 \\ 5 & 4 & 0 \end{pmatrix}^{\textbf{T}}$
, $\mu^{\text{RSSI}}\!\!\! =\!\!\begin{pmatrix} 1 & 3 & 0 \\ 2 & 4 & 0 \end{pmatrix}^{\textbf{T}}$, $\mu^{\text{PDA}} =\begin{pmatrix} 1 & 4 & 3 \\ 5 & 6 & 2 \end{pmatrix}^{\textbf{T}}$. Owing to the proposed utilities, PDA is able to cover all UEs in the match which leads to a higher performance specifically for cell edge users which in this example are UEs $5$ and $6$.\vspace{-0.6cm}
\section{Simulation Results}\label{sec:Simulation Results}\vspace{-0.3cm}
\begin{figure}[!t]
  \begin{center}
   \vspace{-0.2cm}
    \includegraphics[width=7.5cm]{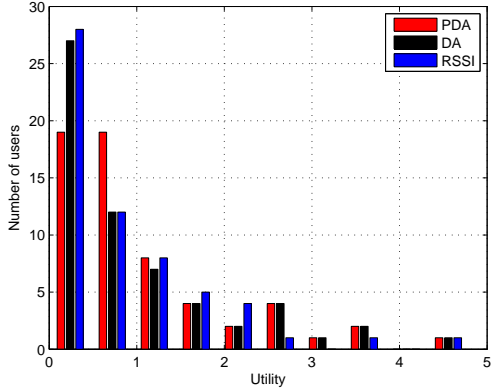}
    \vspace{-0.4cm}
    \caption{\label{fig:hist} The histogram of the utility distribution for different algorithms. Number of users is assumed $M=60$.}
  \end{center}\vspace{-0.8cm}
\end{figure}
 \begin{table}[!t]
\scriptsize
  \centering
  \caption{
    \vspace*{-0.4em}Proposed priority-based deferred acceptance}
    \begin{tabular}{p{8cm}}
      \hline
      \textbf{Inputs:}\,\,$\mathcal{L},\mathcal{M},\mathcal{Q},\textbf{H}$\\
\hspace*{1em}\textit{Initialize:}   \vspace*{.1em}
Calculate the preference matrix \textbf{M} using (5). Initialize temporary rejected vector of users, $\mathcal{R}$.\vspace*{.1em}\\
\textbf{while} $\mathcal{R}$ is nonempty\\
\hspace*{2em}\textit{repeat}\vspace*{.2em}\\
\hspace*{1em}\textit{step 1:} User $m \in \mathcal{R}$ sends its preference vector $\chi_{m}$ to the next BS that is going to apply.\vspace*{.1em}\\
\hspace*{1em}\textit{step 2:} BS $l \in \mathcal{L}$ updates its applicant list, assigns priorities to users as discussed in ~\ref{sec:priority} and calculates the utilities from (6-7). BS $l$ ranks the applicants by their utility and selects first $\mathcal{Q}(l)$ users and rejects the rest.\vspace*{.1em}\\
\hspace*{1em}\textit{step 3:} Acceptance matrix $\mathcal{A}$ and the rejection vector $\mathcal{R}$ get updated. For $\forall m \in \mathcal{R}$:\\
\hspace*{1em}\textit{If C(m)=0},\\
Exclude m from $\mathcal{R}$ and add to unmatched set of users $\mathcal{U}$. \vspace*{.1em}\\
\hspace*{0em}\textbf{Output:}\,\,Stable matching $\mu$\\
   \hline
    \end{tabular}\label{tab:algo}\vspace{-0.6cm}
\end{table}
For simulations, we compare the performance of the proposed matching approach with the RSSI algorithm and the conventional DA proposed in \cite{11}. We consider a total of $10$ SCBSs distributed randomly within a square area
of 1 km $\times$ 1 km with the MBS at the center. The quota per BS is set to a typical value of $4$ UEs~\cite{14}. The channels experience Rayleigh fading, with the propagation loss set to $\alpha_{\text{loss}}=3$. The transmit power of the MBS and the SCBSs are assumed to be 10 W and 1 W, respectively. We assume the noise level to be negligible compared to the interference level. The parameters of the promotion function are set to $\zeta_1=0.1$ and $\zeta_2=3$ and the priority coefficient is set to $\alpha_m \in \{100, 30, 1\}$. Throughout the simulations, the unmatched users are assigned a zero utility. All statistical results are averaged over a large number of independent runs for different locations and channel gains.

In Fig.~\ref{fig:fig2}, we show the average utility per UE resulting from the proposed PDA algorithm and we compare it to both RSSI and DA, as the number of UEs varies.  Fig.~\ref{fig:fig2} shows that, as the number of UEs increases, the average utility of all three schemes decreases due to the quota limitations of each BS. Indeed, the number of unmatched users increases as the total number of users grows. In Fig.~\ref{fig:fig2}, we can see that, at all network sizes, the proposed PDA has a significant advantage in terms of the average utility per UE, reaching up to $65\%$ relative to the RSSI scheme (at $M=80$ UEs). However, in Fig.~\ref{fig:fig2}, we can see that the average utility of DA is comparable to PDA. That is due to the fact that the priorities defined in the PDA provides a fairer allocation between users. In particular, the PDA will allow significant improvements in the worst-case utilities and rates achieved by worst-case SCBSs, as shown in Fig.~\ref{fig:fig3}.

Fig.~\ref{fig:fig3} shows the  SINR-based rates of three approaches for the average rate of worst-case small cell. In Fig.~\ref{fig:fig3}, we can see that as $M$ increases, the average worst-case rates increases. This is a result of the fact that there are more UEs in the proximity of each BS as $M$ increases. This increases the probability of filling up the quotas of the SCBSs with UEs having higher quality links. In this figure, we can clearly see that, at all network sizes, the proposed PDA has a considerable gain compared to DA reaching about $40\%$ of improvement (at $M=70$ UEs) in the worst-case rate. In addition, Fig.~\ref{fig:fig3} shows that the proposed PDA has a comparable worst-case SCBS rate, when compared to the RSSI.

In Fig.~\ref{fig:hist}, we evaluate the performance of the proposed PDA via the utility distribution among UEs for $M=60$ UEs. Fig.~\ref{fig:hist} shows that the proposed PDA has significantly more users achieving higher utilities, when compared to both RSSI and DA. For example, for $M=60$, only $32\%$ of users are assigned to the lowest $10\%$ of utilities, while this value for DA and RSSI is $45\%$ and $47\%$, respectively. This is mainly due to the the fact that the proposed PDA is able to reduce significantly the number of unmatched users.
\begin{figure}[!t]
  \begin{center}
   \vspace{-0.5cm}
    \includegraphics[width=7.5cm]{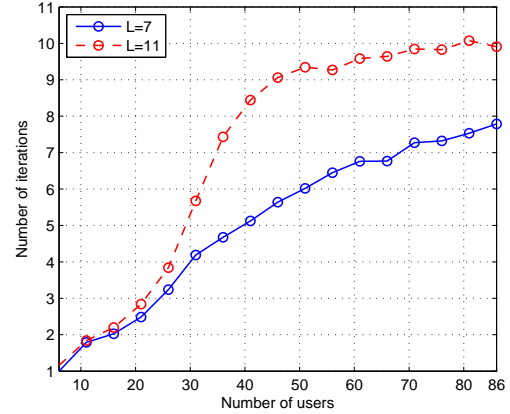}
    \vspace{-0.3cm}
    \caption{\label{fig:it} Average number of iterations versus the number of users.}
  \end{center}\vspace{-1cm}
\end{figure}

Fig.~\ref{fig:it} shows the average number of iterations resulting from the proposed PDA as the number of users $M$ varies, assuming $L=7$ and $L=11$ BSs. In this figure, we can see that, as the number of UEs and SCBSs increase, the average number of iterations increases due to the increase in the number of players. Nonetheless, Fig.~\ref{fig:it} demonstrates that the proposed matching approach has a reasonable convergence time that does not exceed an average of $10$ iterations for a network with $M=86$ users and $10$~SCBSs.\vspace{-0.5cm}
\section{Conclusions}\vspace{-0.3cm}
\label{sec:print}
In this paper, we have proposed a novel approach for cell association in SCNs. We have formulated the problem as a many-to-one matching game in which users and base stations evaluate each other based on well defined utilities. In the proposed utilities, we have introduced a new notion of priorities that allows the base stations to use the information concealed in the preferences of each user in conjunction with conventional rate maximization. We have shown that being aware of each user's overall preferences provides a beneficial insight to the base stations thus allowing an enhanced user association in the downlink of SCNs. To solve the game, we have proposed a self-organizing algorithm that is guaranteed to reach a stable matching.  Simulation results have shown that the proposed approach yields a significant performance improvement in terms of the average utility per user and the average rate experienced by worst-case cells.

\bibliographystyle{IEEEbib}
\bibliography{strings,refs}

\end{document}